\documentclass[published]{JHEP3} 

\JHEP{00(2004)000}

\usepackage{epsfig,multicol}

\newcommand\fverb{\setbox\pippobox=\hbox\bgroup\verb}
\newcommand\fverbdo{\egroup\medskip\noindent%
			\fbox{\unhbox\pippobox}\ }
\newcommand{\be}{\begin{equation}}
\newcommand{\ee}{\end{equation}}
\newcommand\fverbit{\egroup\item[\fbox{\unhbox\pippobox}]}
\newbox\pippobox

\title{SNO, SuperKamiokande data, antineutrinos and sterile neutrinos}

\author{Bhag C. Chauhan \thanks{On leave from Govt. Degree College, 
Karsog (H P) India 171304.}
		and Jo\~{a}o Pulido\\
   Centro de F\'\i sica das Interac\c c\~oes Fundamentais (CFIF) \\
 Departamento de F\'\i sica, Instituto Superior T\'ecnico \\
Av. Rovisco Pais, P-1049-001 Lisboa, Portugal\\
	E-mail: \email{chauhan@cfif.ist.utl.pt}, \email{pulido@cfif.ist.utl.pt
}}
\received{...} 		
\revised{...}
\accepted{...}		

\preprint{\hepph{0406227}}	

\abstract{
Allowing for antineutrinos ($\bar\nu_x$) and sterile neutrinos ($\nu_s$) to accompany 
LMA oscillations, we derive in a model independent way, upper bounds on their 
components in the solar flux, using the recent data from SNO and SuperKamiokande. 
Along with the general case (LMA + $\bar\nu_x$ + $\nu_s$) we consider the special 
cases where only $\bar\nu_x$ or $\nu_s$ are present. We obtain an upper bound 
on $\bar\nu_x$ which is independent of the $\nu_s$ component. In the no sterile 
case we obtain upper and lower bounds on $f_B$, the SSM normalization factor. 
We also investigate in the general case the common parameter range
for $f_B$ and the $\bar\nu_x$, $\nu_s$ components and find
that the upper bound on $\nu_s$ is hardly sensitive to the 
$\bar\nu_x$ component. In the absence of $\bar\nu_x$ we recover the 
$\nu_s$ upper bound existing in the literature. We finally present
a simple $\chi^2$ analysis of all four cases considered.}  

\begin{document} 

\section{Introduction}

Despite the recent realization \cite{Eguchi:2002dm} that the solar neutrino deficit, 
acknowledged over three decades ago \cite{Davis:1968cp}, results mostly from neutrino 
oscillations through LMA \cite{LMA}, it is by no means certain whether oscillations
are accompanied by the conversion in the sun of electron neutrinos into sterile ones,
into antineutrinos of other species, or both. This open question has 
obvious implications in a possible time modulation of the solar neutrino
flux, an effect for which evidence was found by the Stanford Group 
\cite{Sturrock:2004hx},\cite{Sturrock:2003kv}, \cite{Caldwell:2003dw}, 
\cite{Sturrock:2001qn}. In fact, if electron neutrinos
produced in solar fusion reactions interact via a sizeable magnetic moment 
\cite{Lim:1987tk}
with a time varying solar magnetic field, the result is the production of
a time dependent component of active $\bar\nu_{\mu}$ or $\bar\nu_{\tau}$ or 
unobserved sterile neutrinos \cite{Chauhan:2004sf} in the neutrino flux from the
sun, reflecting in some way the time variation of the solar field.

In this article we perform a model independent analysis of the implications from the 
SNO salt phase I and II \cite{Ahmed:2003kj} and SuperKamiokande (SK) results 
\cite{Fukuda:2002pe} on the flux of sterile neutrinos and active antineutrinos which 
may accompany the LMA effect. Being model independent, our analysis will mainly focus 
on solar neutrino data and its implications on bounds of sterile neutrino and 
antineutrino components. Several model independent analyses of solar neutrino data 
have been performed in the past. These concentrated in the survival probability $P_{ee}$ 
\cite{Berezinsky:2001uv} and sterile neutrino component bounds \cite{Barger:2001zs}, 
\cite{Barger:2002iv},\cite{Balantekin:2004hi},\cite{Barger:2001pf},\cite{Cirelli:2004cz}. 
In section 2 we examine the consequences 
from SNO and SK for the joint possibility of active antineutrinos and sterile
neutrinos in the solar flux and for the limiting cases of each of these components
alone. So in this respect our results generalize those of ref. \cite{Balantekin:2004hi}. 
Regarding $\bar\nu_e$, all considerations derived for antineutrinos of 
the other flavours would apply, if not for the recent and very strict   
upper bound on the former from the KamLAND experiment \cite{Eguchi:2003gg}. We find
that SK data on the neutrino electron scattering total rate
leads to the exclusion of all active antineutrinos up to $1.17\sigma$ and
$0.83\sigma$ when combined with SNO data from salt phase I and II respectively. 
Up to $2\sigma$ the percentage of non electron antineutrinos $\bar\nu_x$ 
in the active neutrino flux is smaller than 64\% and 88\% when SK is 
combined with data from SNO I and II respectively. Our results are independent 
of standard solar model normalization. Section 3 deals with a simple $\chi^2$
analysis and in section 4 we expound our main conclusions.

\section{Model Independent Analysis}

We start with the event rate expressions for the charged current (CC) and
neutral current (NC) reactions for SNO and neutrino electron scattering (ES) 
for SK, SNO \cite{Kang:2004tx}
\be 
R^{CC}=f_B P_{ee}
\ee
\be
R^{NC}=f_B P_{ee}+f_{B}(1-P_{ee})[sin{^2}{\alpha}~sin^{2}{\psi}+\bar{r}_d
~sin{^2}{\alpha}~cos^{2}{\psi}]
\ee 
\be
R^{ES}=f_B P_{ee}+f_{B}(1-P_{ee})[r~sin^{2}{\alpha}~sin^{2}{\psi}+\bar r
~sin^{2}{\alpha}~cos^{2}{\psi}].
\ee 
Of all electron neutrinos that are converted, proportional to $1-P_{ee}$,
$sin^{2} \alpha$ denotes the fraction that is converted 
into active ones $\nu_{x},\bar\nu_{x}$, ($x \ne e$) while $\psi$ is 
the angle describing the $\nu_{x},\bar\nu_{x}$ components. The sterile neutrino 
component is therefore proportional to $cos^2{\alpha}$. Parameter
$f_B$ denotes the normalization to the standard solar model
$^8 B$ neutrino flux \cite{Bahcall:2004fg}. Quantities $r$, $\bar r$ are 
respectively the ratios of the NC neutrino and 
antineutrino event rates to the NC+CC neutrino event rate and  
$\bar{r_d}$ is the ratio of the antineutrino deuteron fission to neutrino 
deuteron fission event rate. Specifically
\be
r=\frac{\int dE_{\nu}\phi(E_{\nu})\int dE_{e}\int dE^{'}_{e}\frac{d\sigma_{NC}}
{dE_e}f(E^{'}_e,E_{e})}{\sigma_{NC}\rightarrow \sigma_{NC+CC}}
\ee
\be
\bar r=\frac{\int dE_{\nu}\phi(E_{\nu})\int dE_{e}\int dE^{'}_{e}\frac{d\bar\sigma_{NC}}
{dE_e}f(E^{'}_e,E_{e})}{\bar \sigma_{NC}\rightarrow \sigma_{NC+CC}}
\ee
\be
\bar r_{d}=\frac{\int dE_{\nu}\phi(E_{\nu})\bar\sigma_{NC}(E_{\nu})}
{\bar\sigma_{NC}\rightarrow \sigma_{NC}}
\ee
where $f$ is the energy resolution function for SNO \cite{SNO} or SK \cite{SK}.
Owing to its near energy independence in this range, the electron neutrino survival 
probability $P_{ee}$ is factorized out of these integrals as in eqs.(2.1)-(2.3).  
Energy thresholds considered are $E_{e_{th}}=5.5,~5~MeV$ for SNO and SK respectively 
and the rest of the notations is standard. We obtain
\be
r=0.150~({\rm 0.151~for~SK}),~\bar{r}=0.115~({\rm~0.116~for~SK}),~\bar{r_d}=0.954.
\ee
where these minor differences are mainly the result of the difference in the threshold 
energies, being largely independent of the resolution functions. The data from SNO
(phase I and II) and SK are summarized in table I.

\begin{center}
\begin{tabular}{cccc} \\ \hline \hline
& CC & NC & ES \\ \hline
SNO I & $0.275\pm^{0.017}_{0.018}$ & $0.900\pm 0.081$ & $0.382\pm^{0.056}_{0.048}$ \\
SNO II & $0.294\pm^{0.020}_{0.021}$ & $0.846\pm^{0.065}_{0.062}$ & $0.368\pm^{0.056}_{0.050}$ \\
SK     & & & $0.406\pm^{0.013}_{0.011}$ \\ \hline
\end{tabular}
\end{center}

{\it{Table I - Ratios of event rates to standard solar model \cite{Bahcall:2004fg} event rates at SNO 
and SK (theoretical error not considered).}} 

From eqs.(2.1)-(2.3) one can eliminate the angle $\alpha$ and express angle $\psi$ in
terms of the experimental event rates and other model independent quantities,
\be
sin^2 \psi= \frac{\bar{r} -\gamma \bar{r_d}}{\gamma (1-\bar{r_d})+\bar{r}-r}
\ee
with 
\be
\gamma=\frac{R^{ES}-R^{CC}}{R^{NC}-R^{CC}}.
\ee 

If not for the large uncertainties that are propagated into equation (8) originated from 
the uncertainties in $R^{ES},~R^{CC},~R^{NC}$, this would unambiguously determine the
relative proportion of non electron antineutrinos $\bar\nu_x$ in the active non $\nu_e$ 
flux. Hence, as will be seen, only upper bounds on the fraction of the $\bar\nu_{x}$ 
flux can be derived at the present stage. To this end we evaluate the parameter 
$sin^2 \psi$ using eq. (2.8) for all values of $R^{ES},~R^{CC},~R^{NC}$ within their 
allowed $1\sigma$ ranges for SNO: these are represented by the light shaded areas in fig.1 
where the SNO data used are those from salt phase II. Hence for each chosen value of 
$sin^2 \psi$ the allowed values of the three reduced rates lie within each shaded
area. If the SNO experiment alone 
is considered, it is seen that all possible values of $sin^2 \psi$ in the range 
$0\leq sin^2 \psi\leq 1$ can be obtained. We note that as the $\bar\nu_{x}$ 
component decreases and eventually vanishes ($sin^2 \psi \rightarrow 1$), the factor 
multiplying $r$
in eq.(2.3) increases while the one multiplying $\bar r$ approaches zero. Owing to
the relative difference between $r$ and $\bar r$ [eq.(2.7)], this implies a slight 
inclination into larger values of $R^{ES}$ for a decreasing $\bar\nu_{x}$ component. 
The same effect, although much less significant, because of the much smaller 
difference between $\bar r_d$ and unity, is also present in the ($sin^2 \psi,R^{NC}$) 
area and does not exist in $R^{CC}$ [see eq. (2.1)]. 

The data from the SK experiment with $R^{ES}$ restricted to its SK $1 \sigma$ range 
(see table I) are also used to evaluate $sin^2 \psi$. In fig.1 the dark shaded area 
which is part of the total ($sin^2 \psi,R^{ES}$) one represents the parameter range allowed
jointly by SNO II and SK. This is enlarged in fig.2 with a magnified horizontal
scale. The result is a lower bound on $sin^2 \psi$
\be
{\rm SNO~II}~~sin^2 \psi >0.5~{\rm at}~1.49\sigma(87\%~CL)
\ee         
which is in fact an upper bound (0.5) on the fraction of non-electron antineutrinos 
$\bar\nu_{X}$ in the active non-$\nu_e$ flux\footnote{Recall that the $\bar\nu_{X}$
component is proportional to $cos^2 \psi$ [see eqs.(2.2), (2.3)].}.
The same procedure, as applied to SNO I with SK leads to
\be
{\rm SNO~I}~~sin^2 \psi >0.95~{\rm at}~1.23\sigma(79\%~CL).
\ee
The upper bound on $\bar\nu_{x}$ is therefore more restrictive (0.05) if one considers 
SNO I data. Up to 95\% CL these bounds become
\be
sin^2 \psi >0.12~({\rm SNO~II})~,~sin^2 \psi >0.36~({\rm SNO~I})~~(95\%~CL)
\ee
hence respectively an upper bound of 0.88 and 0.64 on the $\bar\nu_{x}$ fraction. 
These results are independent of the normalization to any particular standard solar 
model. No restriction on the sterile neutrino component, proportional to $cos^2 \alpha$ 
[see eqs.(2.2), (2.3)], has been considered so far, therefore our analysis is valid for any 
$\nu_s$, $\bar \nu_x$ admixture accompanying the LMA effect. Combining separately 
eqs.(2.1), (2.2) and eqs.(2.1), (2.3) one can relate the normalization factor $f_B$ to the
mixing angles $\alpha$, $\psi$. We have respectively
\be
f_B=R^{CC}+\frac{R^{NC}-R^{CC}}{sin^2 \alpha (sin^2 \psi +\bar r_{d}~cos^2 \psi)}
\ee 
\be
f_B=R^{CC}+\frac{R^{ES}-R^{CC}}{sin^2 \alpha (r~sin^2 \psi +\bar r~cos^2 \psi)}.
\ee 

We represent in fig.3 the allowed range of $sin^2 \alpha$ (proportional to the
active non-$\nu_e$ component) in terms of $f_B$ at the 95\% CL for SNO II 
(fig.3a) and SNO I (fig.3b) using inequalities (2.12) and eq.(2.13).
Considering the two possible equivalent choices to generate fig.3, namely
eqs.(2.13) and (2.14), the former should in fact be preferred since it leads to 
the narrowest error bars in $f_B$. The dashed and full lines in fig.3 correspond 
respectively to $sin^2 \psi=1$ (no antineutrinos $\bar\nu_{x}$) and to the
95\% CL upper limits for $\bar\nu_{x}$. These are $sin^2 \psi=0.12$ for SNO II 
and $sin^2 \psi=0.36$ for SNO I [see eq.(2.12)]. Hence the existing 
shift between each adjacent dashed and full line represents the small change in 
the sterile neutrino component, proportional to $cos^2 \alpha$, resulting from 
introducing in the scheme a $\bar\nu_{X}$ component up to its 95\% CL upper
bound. This shows 
that the possible sterile neutrino flux is hardly sensitive to the presence of 
antineutrinos, a fact whose origin becomes clear on examination of the denominator in 
eq.(2.13): the multiplier of $sin^2 \alpha$ is very close to unity for any value 
of $\psi$ owing to the fact that $\bar r_{d} \simeq 1$. 
From fig.3 it is also seen that in the absence of $\bar\nu_x$ ($x \ne e$, 
$sin^2 \psi=1$) the fraction of solar neutrinos oscillating to active ones 
is greater than 0.59 (SNO II) and 0.63 (SNO I) at $2 \sigma$ of the non-$\nu_e$ flux.  
Allowing 
for non-electron antineutrinos up to their $2 \sigma$ upper bound this fraction 
becomes respectively 0.62 and 0.66. This result is consistent with the result of 
ref. \cite{Balantekin:2004hi} where the authors also included KamLAND data 
in their analysis but were restricted to the case $sin^2 \psi=1$.    

We now take an alternative view by considering separately the cases in which 
either only $\bar \nu_x$ or $\nu_s$ 
is present along with LMA and derive in each the corresponding constraints on the 
SSM normalization factor $f_B$. We start with the case where no steriles are present 
(only $\bar \nu_x$). Here $sin^2 \alpha=1$ and from eqs.(2.2), (2.3) one obtains
\be
f_B=R^{CC}+\frac{(R^{NC}-R^{CC})(r-\bar r)-(R^{ES}-R^{CC})(1-\bar r_{d})}
{\bar r_{d}(r-\bar r)-\bar r(1-\bar r_{d})}
\ee
which for SNO II and SNO I give respectively, using table I (to 1$\sigma$)
\be
f_B=0.86\pm0.12~{\rm (SNO~II)},~~f_B=0.88\pm0.13~{\rm (SNO~I)}.
\ee
We note that these correspond to the allowed ranges within the lines $sin^2 \alpha=1$ 
in the two panels of fig.3, the slight discrepancies with this figure 
being of course the result of the experimental uncertainties and the different 
procedures used for generating the two sets of results.
For the combined SNO and SK data, eq.(2.16) becomes instead (to 1$\sigma$)
\be
f_B=0.80\pm0.09~{\rm (SNO~II+SK)},~~f_B=0.84\pm0.10~{\rm (SNO~I+SK)},
\ee
the smaller error resulting from the smaller SK error. All these parameter ranges
lie within the allowed $1 \sigma$ SSM error of 23\% \cite{Bahcall:2004fg}. It is
thus seen that the former general analysis which includes antineutrinos and steriles, 
and whose results are summarized in fig.3, leads to more precise predictions for $f_B$,
as only two experimentally measured quantities $R^{NC}$ and $R^{CC}$ are used in 
contrast to eq.(2.15). In fact, in fig.3, where all quantities are allowed to vary 
within their $2\sigma$ ranges, we have (for $sin^2 \alpha=1$) 
\be
f_B=0.87\pm0.15~{\rm (SNO~II)},~~f_B=0.91\pm0.19~{\rm (SNO~I)}
\ee
to be compared with eq.(2.16) where only $1\sigma$ ranges are allowed.

We now briefly refer to the other special case, namely the absence 
of antineutrinos: only steriles are present here along with the LMA effect, hence 
$sin^2 \psi=1$. This case corresponds to the shaded areas in fig.3 limited by the
two dashed lines and, in contrast to the previous one,  
no model independent equation can be obtained for $f_B$, but only a degeneracy 
relation between $f_B$ and $sin^2 \alpha$. This can be expressed by either of the two
equivalent equations
\be
f_B=R^{CC}+\frac{R^{NC}-R^{CC}}{sin^2 \alpha}
\ee
\be
f_B=R^{CC}+\frac{R^{ES}-R^{CC}}{r~sin^2 \alpha}
\ee
which correspond to eqs.(2.13) and (2.14) with $sin^2 \psi=1$. As previously discussed
in the general case (LMA + $\bar\nu_x$ + $\nu_s$) the main result here is an upper 
bound on the sterile component. At $2\sigma$ this is $cos^2\alpha < 0.41$ (from SNO II) 
or 0.38 (from SNO I) of the non-$\nu_e$ flux for $f_B=1$. 

\section{Introducing $\chi^2$ Analysis}

We refine our results by performing a $\chi^2$ analysis of all four cases considered.
The $\chi^2$ definition is quite simple \cite{Barger:2001pf}
\begin{equation}
\chi^2=\sum_{i} \frac{(R_i-R_{i}^{th})^2}{\delta R_{i}^2}
\end{equation}
where the sum extends over the four experiments $(i=ES_{SK},ES_{SNO},NC,CC)$,
$R_i$, $\delta R_{i}$ denote the experimental reduced rates and their errors quoted
in table I, and $R_{i}^{th}$ are given by eqs. (2.1)-(2.3). The result of the
$\chi^2$ minimization is shown in tables II, III for SNO II and SNO I respectively.

\newpage
\begin{center}
\begin{tabular}{cccccc} \\ \hline \hline
 & $f_B$ & $P_{ee}$ & $sin^2 \alpha$ & $sin^2 \psi$ & $\chi^2_{min}$ \\ \hline
LMA (2 dof) & 0.876 & 0.356 & 1.0 & 1.0 & 1.67 \\
LMA+$\bar\nu_{x}$ (1 dof) & 0.876 & 0.356 & 1.0 & 1.0 & 1.67 \\
LMA+$\nu_{s}$ (1 dof) & 0.961 & 0.324 & 0.869 & 1.0 & 1.67 \\
LMA+$\bar\nu_{x}+\nu_{s}$ (0 dof) & 0.989 & 0.315 & 0.833 & 1.0 & 1.67 \\ \hline
\end{tabular}
\end{center}
 
{\it{Table II - Results of $\chi^2$ analysis for SNO II.}}

\vspace{0.8cm}

Inspection of table II (second row) shows that the best fit for case LMA+$\bar \nu_x$
corresponds to the very absence of $\bar \nu_x$ ($sin^2 \psi=1$). It is also seen
that allowing for $\nu_s$ alone in addition to LMA (third row) as well as 
LMA+$\bar \nu_x+\nu_s$ (fourth row) leads to a best fit solution with a small
although non negligible $\nu_s$ component (13\% and 17\% respectively).
Furthermore table II also shows that $\chi^2_{min}$ 
is independent of the values of $f_B,~P_{ee},~
sin^2 \alpha$. However it depends on $sin^2 \psi$: if in fact we let
$sin^2 \psi$ to be unconstrained, an absolute $\chi^2_{min}$ is obtained
for an unphysical value of $sin^2 \psi$,
\footnote{These are $\chi^2_{min}=0.384$, $sin^2 \psi=2.84$ (SNO II) and
$\chi^2_{min}=0.133$, $sin^2 \psi=3.17$ (SNO I).}. As long as $sin^2 \psi$ remains
constrained to its physical region (0 $\leq sin^2 \psi \leq 1$), $\chi^2_{min}$
is fixed regardless of the values of the other three parameters $f_B,~P_{ee},~
sin^2 \alpha$. A similar situation is observed in SNO I (see table III) with
the sterile component totally missing ($sin^2 \alpha=1$) in the LMA+$\nu_s$
case. This reflects the fact that the parameters $f_B,~P_{ee},~sin^2 \alpha$ 
can be eliminated from eqs. (2.1) - (2.3) so as to express 
$sin^2 \psi$ ($\bar \nu_x$ component) in terms of experimentally measured 
quantities only (see eqs. (2.8), (2.9)). Likewise the bounds on $sin^2 \psi$
are independent of the $\nu_s$ component and of the other two parameters
$f_B,~P_{ee}$ (see eqs. (2.10), (2.12)).

\begin{center}
\begin{tabular}{cccccc} \\ \hline \hline
 & $f_B$ & $P_{ee}$ & $sin^2 \alpha$ & $sin^2 \psi$ & $\chi^2_{min}$ \\ \hline
LMA (2 dof) & 0.965 & 0.304 & 1.0 & 1.0 & 2.47 \\
LMA+$\bar\nu_{x}$ (1 dof) & 0.965 & 0.304 & 1.0 & 1.0 & 2.47 \\
LMA+$\nu_{s}$ (1 dof) & 0.965 & 0.304 & 1.0 & 1.0 & 2.47 \\
LMA+$\bar\nu_{x}+\nu_{s}$ (0 dof) & 0.969 & 0.302 & 0.933 & 1.0 & 2.47 \\ \hline
\end{tabular}
\end{center}
{\it{Table III - Results of $\chi^2$ analysis for SNO I.}}

\vspace{0.8cm}
Finally, the 2$\sigma$ upper bound on the sterile component mentioned at the 
end of section 2 is also shown in the contour plots of fig.4 and
corresponds to the lower edge of the light shaded area. 

\section{Conclusions}

To conclude, the results of this paper can be summarized in figs.2, 3 and 
eqs. (2.16), (2.17), (2.18). Allowing for antineutrinos $\bar\nu_x$ other than 
$\bar\nu_e$ and sterile neutrinos, both possibly generated in the sun through spin 
flavour precession accompanying LMA, we have derived in a model independent way, 
using SNO and SK data, upper bounds on the flux of these solar antineutrinos and 
steriles. We related these bounds to the parameter ranges allowed for the SSM 
normalization factor $f_B$. To summarize:
 
{\it (i)} We found an upper bound for $\bar\nu_x$ which at $2\sigma$ is 0.88 (SNO II) 
or 0.64 (SNO I) of the active non-$\nu_e$ flux [see fig.2 and eq.(2.12)]. This is 
independent of the sterile neutrino component. 

{\it (ii)} In the no sterile case we obtained upper and lower bounds on $f_B$ 
[eqs.(2.16), (2.17), (2.18)]. 

{\it (iii)} In the no $\bar\nu_x$ case (only steriles accompanying LMA) the fraction 
of solar neutrinos oscillating to active ones was found to be  greater than
0.59 (SNO II) or 0.63 (SNO I) of the non-$\nu_e$ flux, a result consistent 
with ref. \cite{Balantekin:2004hi} which is in fact an upper bound on $\nu_s$. 

{\it (iv)} Allowing, in the preceding situation, for $\bar\nu_x$ up to its 
$2\sigma$ upper bound, these limits are increased by only 5\%, (decrease on 
$\nu_s$ upper bound) which shows how the possible $\nu_s$ flux is hardly sensitive 
to the $\bar\nu_x$ component.  

{\it (v)} $\chi^2$ analysis shows that the most disfavoured case (if not excluded) 
is $\bar \nu_x$ either with LMA or with LMA+$\nu_s$. In SNO II it is seen that 
some possibility is left for LMA+$\nu_s$.

\vspace{0.8cm}

{\bf Acknowledgements} \\
{\em  The work of BCC  was supported by Funda\c{c}\~{a}o para a 
Ci\^{e}ncia e a Tecnologia through the grant SFRH /BPD/5719/2001.}


\newpage

\begin{figure}[h]
\setlength{\unitlength}{1cm}
\begin{center}
\hspace*{-1.6cm}
\epsfig{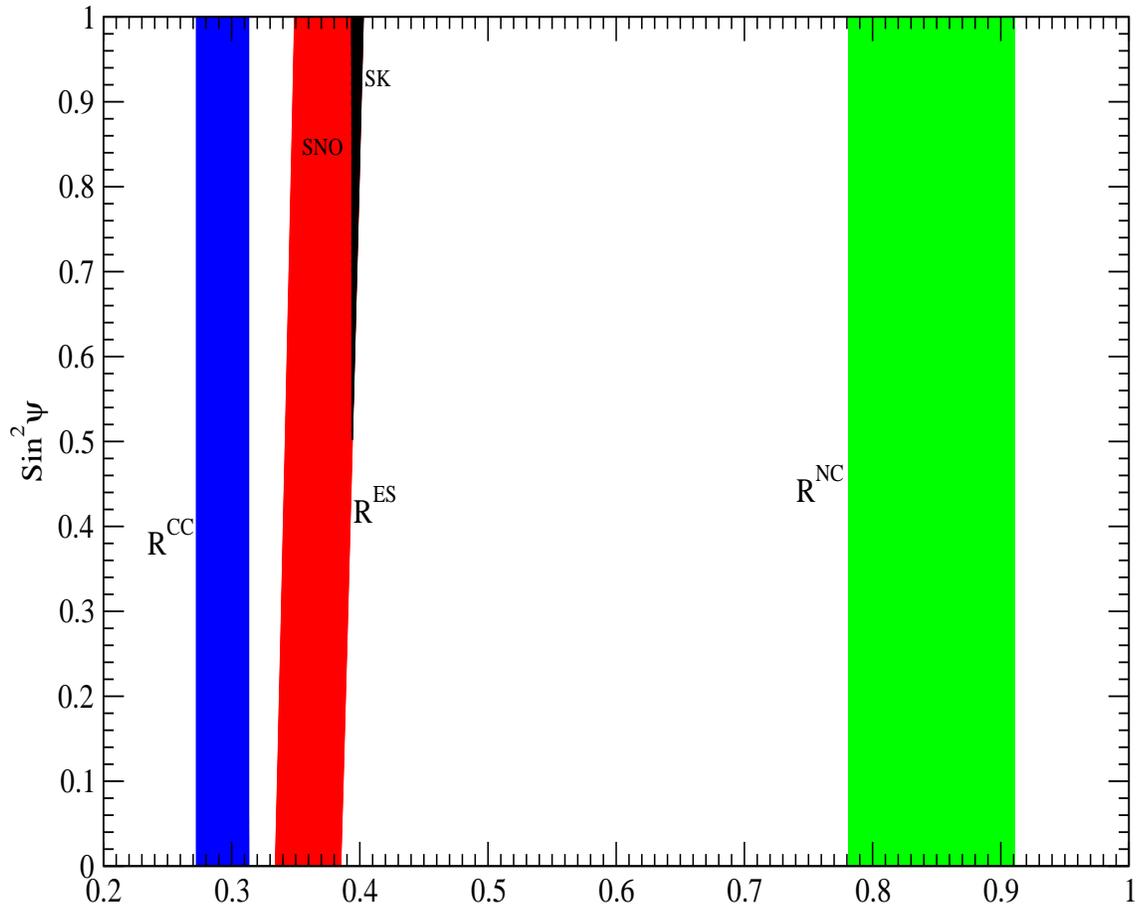}
\end{center}
\caption{ \it The coloured areas are the regions allowed by the $1\sigma$  
ranges of the reduced rates, as reported by the SNO II experiment, and $sin^2 \psi$,
proportional to the neutrino component of the active non-$\nu_e$
flux. The dark shaded area is the region allowed jointly by 
the SNO II and SK data on the electron scattering reduced rate.} 
\label{fig1}
\end{figure}

\begin{figure}[h]
\setlength{\unitlength}{1cm}
\begin{center}
\hspace*{-1.6cm}
\epsfig{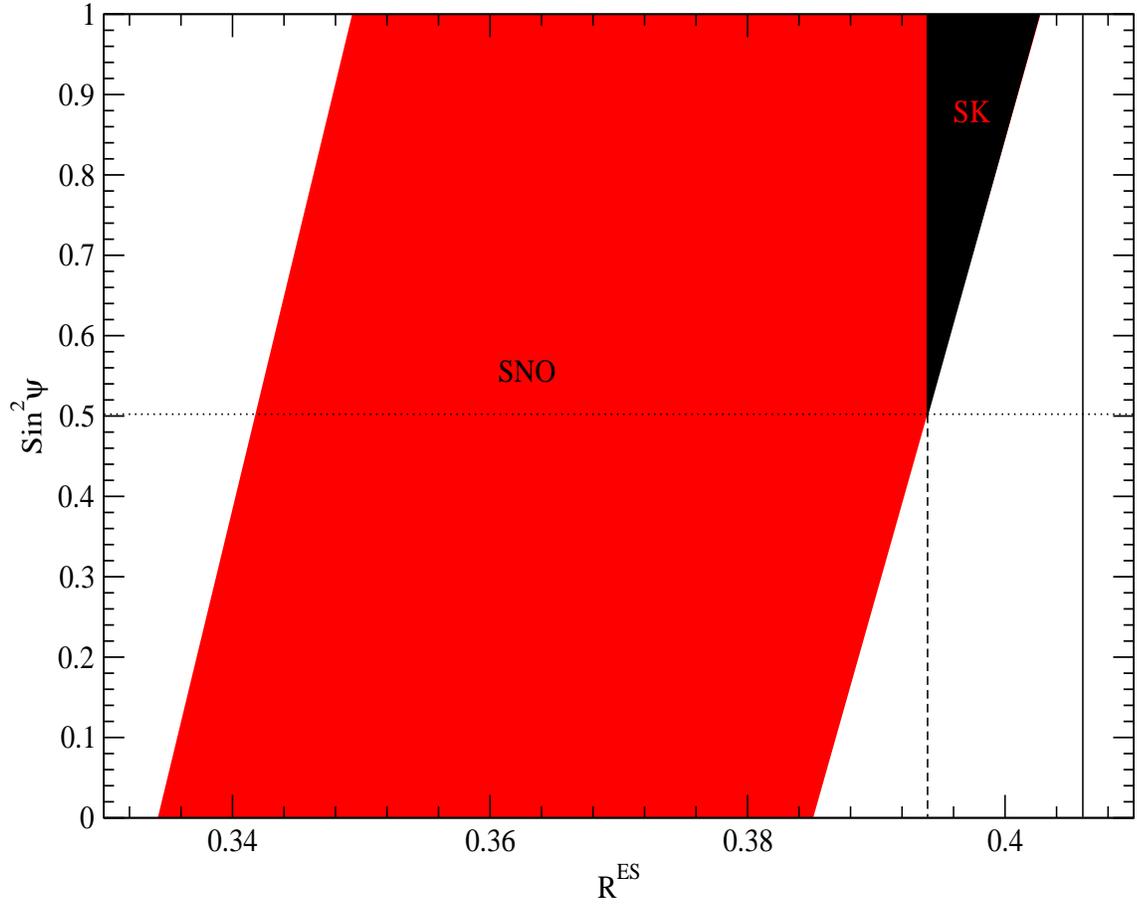}
\end{center}
\caption{ \it Same as fig.1 with a magnified horizontal scale to show the lower
bound on $sin^2 \psi$ implied by the data from the two experiments. This bound is
independent from the sterile neutrino component (see also the main text).}   
\label{fig2}
\end{figure}

\begin{figure}[h]
\setlength{\unitlength}{1cm}
\begin{center}
\hspace*{-1.6cm}
\epsfig{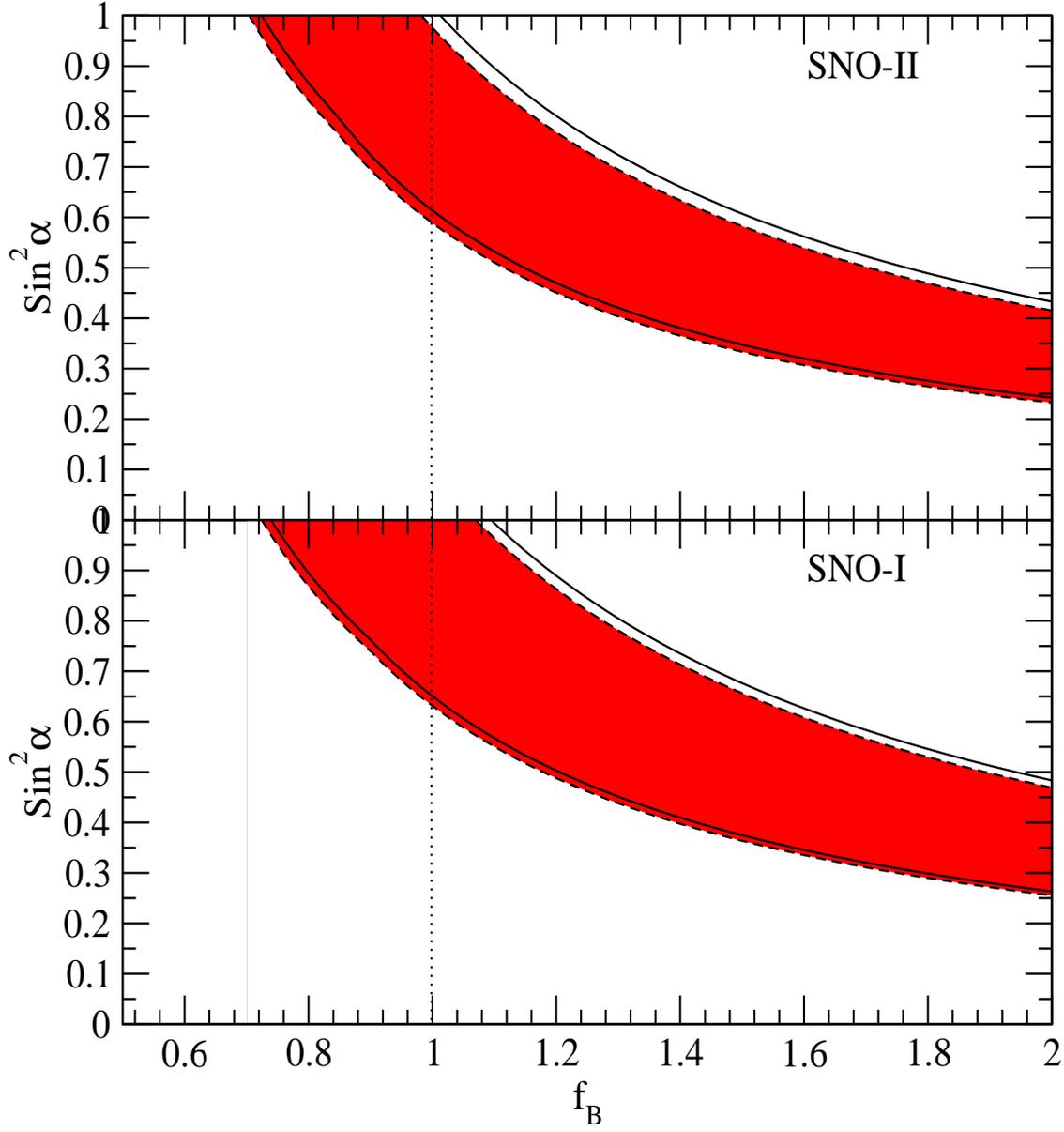}
\end{center}
\caption{ \it The allowed range at $2\sigma$ of $sin^2 \alpha$ (proportional
to the total active non-$\nu_e$ flux) and $f_B$, the SSM normalization factor, using
SNO II and SNO I data. Dashed lines correspond to absence of antineutrinos, 
while full lines to their upper bound at $2\sigma$. It is seen that the
sterile component, proportional to $cos^2 \alpha$, is hardly affected by   
the presence of antineutrinos.}   
\label{fig3}
\end{figure}

\begin{figure}[h]
\setlength{\unitlength}{1cm}
\begin{center}
\hspace*{-1.6cm}
\epsfig{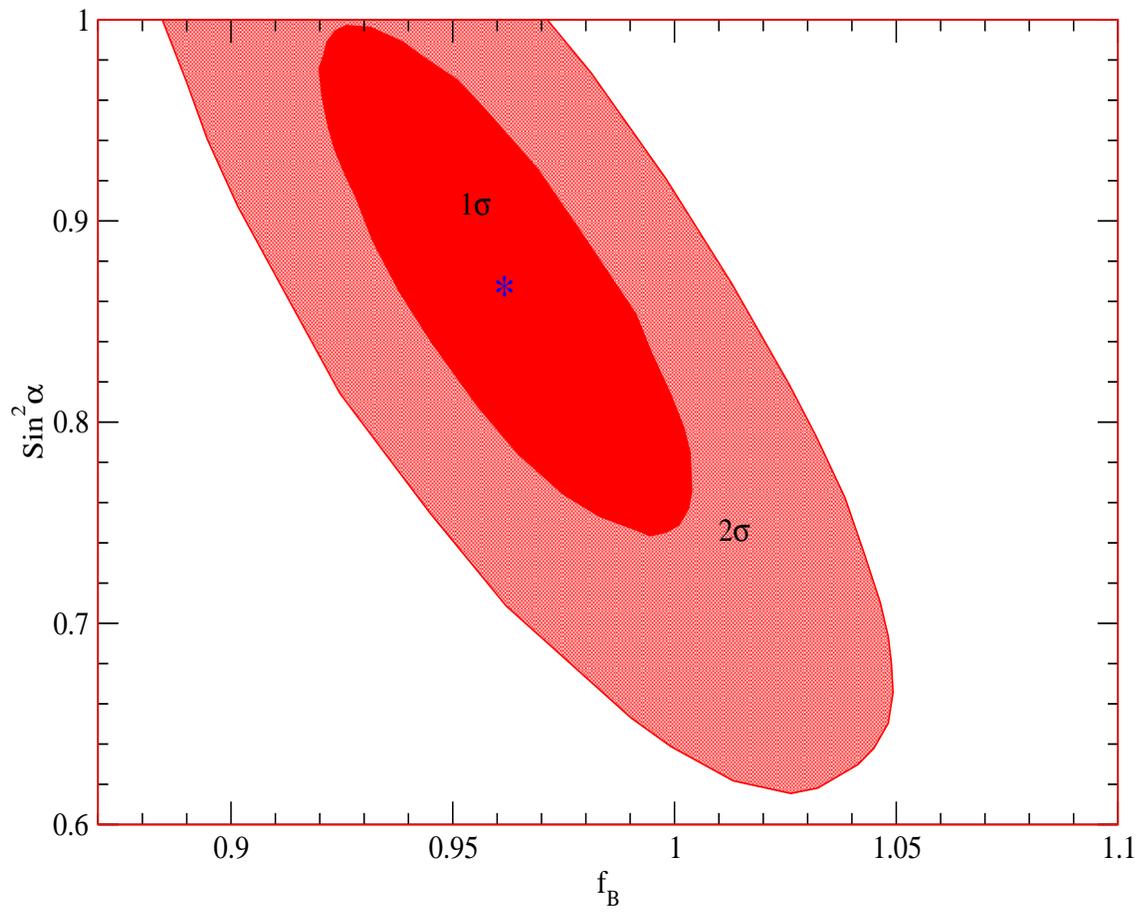}
\end{center}
\caption{ \it $1 \sigma$ and $2 \sigma$ contour plots in the $f_B$, 
$sin^2 \alpha$ plane for SNO II data. The star represents the best fit point.}   
\label{fig4}
\end{figure}

\end{document}